\documentclass[letterpaper]{sig-alternate}

\newfont{\mycrnotice}{ptmr8t at 7pt}
\newfont{\myconfname}{ptmri8t at 7pt}
\let\crnotice\mycrnotice%
\let\confname\myconfname%

\permission{Permission to make digital or hard copies of all or part of this work for personal or classroom use is granted without fee provided that copies are not made or distributed for profit or commercial advantage and that copies bear this notice and the full citation on the first page. \\}
\conferenceinfo{UCL Technical Report}{23 July, 2014, London, UK.}
\copyrightetc{Copyright 2014 University College London}

\usepackage{alex}
\usepackage{url}
\usepackage{epstopdf}
\usepackage{multicol}
\usepackage{subfigure}
\usepackage{algorithm}
\usepackage{algorithmic}

\title{Real-Time Bidding Benchmarking with iPinYou Dataset}
\numberofauthors{2}
\author{
\alignauthor Weinan Zhang, Shuai Yuan, Jun Wang\\
       \affaddr{University College London}\\
       \email{\{w.zhang, s.yuan, j.wang\}@cs.ucl.ac.uk}
\alignauthor Xuehua Shen\\
       \affaddr{iPinYou Inc.}\\
       \email{x@ipinyou.com}
}

\begin{document}

\maketitle

\begin{abstract}
Being an emerging paradigm for display advertising, Real-Time Bidding (RTB) drives the focus of the bidding strategy from context to users' interest by computing a bid for each impression in real time.
The data mining work and particularly the bidding strategy development becomes crucial in this performance-driven business. However, researchers in computational advertising area have been suffering from lack of publicly available benchmark datasets, which are essential to compare different algorithms and systems.
Fortunately, a leading Chinese advertising technology company iPinYou decided to release the dataset used in its global RTB algorithm competition in 2013.
The dataset includes logs of ad auctions, bids, impressions, clicks, and final conversions. These logs reflect the market environment as well as form a complete path of users' responses from advertisers' perspective. This dataset directly supports the experiments of some important research problems such as bid optimisation and CTR estimation. To the best of our knowledge, this is the first publicly available dataset on RTB display advertising. Thus, they are valuable for reproducible research and understanding the whole RTB ecosystem.
In this paper, we first provide the detailed statistical analysis of this dataset. Then we introduce the research problem of bid optimisation in RTB and the simple yet comprehensive evaluation protocol. Besides, a series of benchmark experiments are also conducted, including both click-through rate (CTR) estimation and bid optimisation.
\end{abstract}



\section{Introduction}
\begin{figure}
 \centering
 \includegraphics[width=\columnwidth]{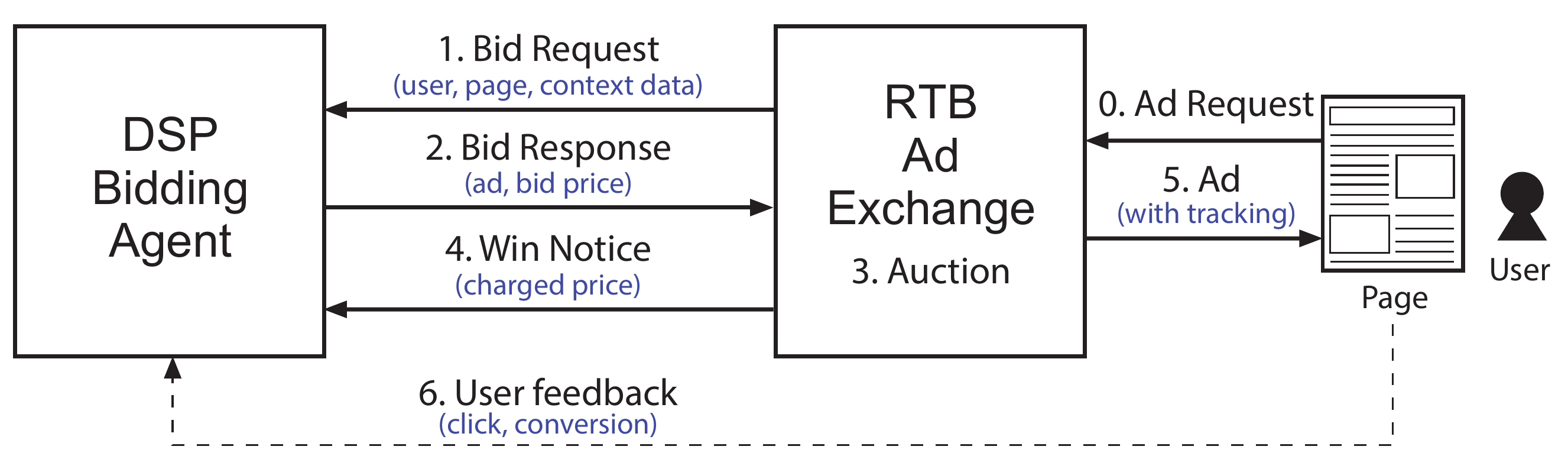}
 \caption{A brief illustration of the interactions between user, ad exchange and DSP.}
 \label{fig:interaction}
\end{figure}

Emerged in 2009 \cite{yuan2013real}, Real-Time Bidding (RTB) has become an important new paradigm in display advertising \cite{muthukrishnan2009ad,chakraborty2010selective,google2011arrival}.
For example, eMarketer estimates a 73\% spending growth on RTB in United States during 2013, which accounts for 19\% of the total spending in display advertising \cite{emarketer2013rtb}.
Different from the conventional negotiation or pre-setting a fixed bid for each campaign or keyword, RTB enables the advertisers to give a bid for every individual impression.
A concise interaction process between the main components of RTB ecosystem is shown in Figure~\ref{fig:interaction}.
Each ad placement will trigger an auction when the user visits an ad-supported site (e.g., web page, streaming videos and mobile apps). Bid requests will be sent via the ad exchange to the advertisers' buying systems, usually referred to as Demand-Side Platforms (DSPs).
Upon receiving a bid request, a DSP will calculate a bid as the response after holding an internal auction among all of its qualifying campaigns.
An auction will be held at each intermediary (ad networks, ad exchanges, etc.) and finally in the publishers' system.
Finally, the winner's ad will be shown to the visitor along with the regular content of the website.
It is commonly known that a long time page-loading would greatly reduce users' satisfactory \cite{muthukrishnan2009ad}, thus, DSPs are usually required to return a bid in a very short time frame (e.g. 100 ms). More detailed introductions to RTB could be found in \cite{Wang:RTB:Tutorial,yuan2013real}.

Algorithms employed by DSPs are expected to contribute a much higher return-on-investment (ROI) comparing with the traditional channels. It is crucial that such algorithms can quickly decide whether and how much to bid for a specific impression, given the contextual and behaviour data (usually referred to as user segments). This is apparently also an engineering challenge considering the billion-level bid requests that a DSP could normally see in a day.

Despite its popularity, the majority research activities on RTB have been limited in advertising technology companies \cite{muthukrishnan2009ad,lee2012estimating,perlich2012bid} so far. It is nearly impossible for researchers from academia to access the sensitive thus highly protected data.
Fortunately, a three-season global competition of RTB algorithms was held by iPinYou\footnote{\url{http://www.ipinyou.com.cn}} in 2013. As will be discussed in Section \ref{sec:task}, the competition task focuses on the bidding strategies from the DSP's perspective: it aims to maximise the campaign's Key-Performance-Indicator (KPI) with the budget and lifetime constraint by developing the bidding strategy. We refer such task as \emph{DSP Bid Optimisation} problem. In March 2014, the dataset used in the three seasons of the competition (about 35 GB) was released for the purpose of research.
To the best of our knowledge, this is the first large-scale real-world RTB dataset. We believe it will stimulate the interest of RTB research and development of DSP bidding algorithms in the whole data science research community, and further speed up the growth of RTB display advertising ecosystem. The dataset can be directly downloaded from our website of computational advertising research\footnote{\url{http://data.computational-advertising.org}}.

In this paper, we first report a detailed statistical analysis of this dataset. Then, we formally present the research problem of DSP bid optimisation and its simple yet comprehensive evaluation protocol. Finally, we show the experimental results of some benchmark bidding strategies as well as the click-through rate (CTR) estimation models.

\section{The iPinYou RTB Dataset}\label{sec:dataset}
\subsection{iPinYou Demand-Side Platform}
iPinYou Information Technologies Co., Ltd (iPinYou) was founded in 2008 and is currently the largest DSP in China. iPinYou is headquartered in Beijing and has offices in Shanghai, Guangzhou and Silicon Valley. iPinYou has built world class RTB technology and algorithm, proprietary cloud computing platform and patented audience profiling technology. It has served over 1000 brands in IT, financial service, auto, consumer packaged goods, travel, electric commerce, gaming and more. It has also significantly improved the advertising effectiveness and fostered extensive partnerships with domestic mainstream media and private exchanges. It is established as a leading provider of audience based programmatic advertising technology.

\subsection{Data Format}
There are four different types of logs in the iPinYou dataset: bids, impressions, clicks, and conversions. The logs are organised on a row-per-record basis.

The feature description and example of each column of the ad log data are presented in Table~\ref{tab:data-format}. Generally, each record contains three kinds of information: (i) The auction and ad features (all columns except 3, 20 and 21). These features are sent to the bidding engine to make a bid response. (ii) The auction winning price (column 21), i.e. the highest bid from the competitors. If the bidding engine responses a bid higher than the auction winning price, the DSP will win this auction and get the ad impression. (iii) The user feedback (click and conversion) on the ad impression (column 3). If the DSP wins the auction, the user feedback on this ad impression can be checked to update the DSP performance.

Note that all numbers related to money (e.g., bid price, paying price and floor price) use the currency of RMB and the unit of Chinese fen $\times 1000$, corresponding to the commonly adopted cost-per-mille (CPM) pricing model. However, in our analysis the calculated numbers (e.g., cost, average CPM, and effective cost-per-click) are not multiplied by 1000.

\begin{table}[t]
\center
\caption{The log data format. Columns with $*$ are hashed or modified before releasing; columns with $\dag$ are only available in impression/click/conversion logs but not in bid logs. }
\label{tab:data-format}
\begin{tabular}{rll}
\small
Col \# & Description & Example \\ \hline
\\ [-2.0ex] 
$^*$1 & Bid ID & 015300008...3f5a4f5121\\ 
2 & Timestamp & 20130218001203638\\
$^\dag$3 & Log type & 1\\
$^*$4 & iPinYou ID & 35605620124122340227135\\
5 & User-Agent & Mozilla/5.0 (compatible; \textbackslash \\
 & & MSIE 9.0; Windows NT \textbackslash \\
 & & 6.1; WOW64; Trident/5.0)\\
$^*$6 & IP & 118.81.189.*\\
7 & Region & 15\\
8 & City & 16\\
$^*$9 & Ad exchange & 2\\
$^*$10 & Domain & e80f4ec7...c01cd1a049\\ 
$^*$11 & URL & hz55b00000...3d6f275121\\ 
12 & Anonymous URL ID & Null\\
13 & Ad slot ID & 2147689\_8764813\\
14 & Ad slot width & 300\\
15 & Ad slot height & 250\\
16 & Ad slot visibility & SecondView\\
17 & Ad slot format & Fixed\\
$^*$18 & Ad slot floor price & 0 \\
19 & Creative ID & e39e178ffd...1ee56bcd\\ 
$^*$20 & Bidding price & 753\\
$^{*\dag}$21 & Paying price & 15\\
$^{*\dag}$22 & Key page URL & a8be178ffd...1ee56bcd\\ 
$^{*}$23 & Advertiser ID & 2345\\
$^{*}$24 & User Tags & 123,5678,3456\\
\end{tabular}
\end{table}

Along with Table~\ref{tab:data-format} we want to give more detailed description for some of the columns here: \\
\textsc{(c01)} The bid ID serves as the unique identifier of all event logs and could be used to join bids, impressions, clicks, and conversions together. \\
\textsc{(c02)} The column uses the format of yyyyMMddHHmmssSSS\footnote{\url{http://docs.oracle.com/javase/8/docs/api/java/time/format/DateTimeFormatter.html#patterns}}.\\
\textsc{(c03)} The possible values include: 1 (impression), 2 (click), and 3 (conversion). \\
\textsc{(c04)} The internal user ID set by iPinYou. \\
\textsc{(c05)} The column describes the device, operation system, and browser of the user. \\
\textsc{(c10)} The domain of the hosting webpage of the ad slot. The values were hashed. \\
\textsc{(c11)} The URL of the hosting webpage of the ad slot. The values were hashed. \\
\textsc{(c12)} When URL is not directly available to the DSP (e.g. masked by ad exchanges) this column will be used. The values are provided by ad exchanges. For one record, either URL or Anonymous URL ID is meaningful. \\
\textsc{(c16)} The column describes if the ad slot is above the fold (``FirstView'') or not (``SecondView'' to ``TenthView''), or unknown (``Na''). \\
\textsc{(c17)} Possible values include ``Fixed'' (fixed size and position), ``Pop'' (the pop-up window), ``Background", ``Float", and ``Na" which presents unknown cases. \\
\textsc{(c18)} Floor (or reserve) price of the ad slot. No bid lower than the floor price could win auctions. A linear scale normalisation was applied to this column. \\
\textsc{(c20)} The bid price from iPinYou for this bid request. \\
\textsc{(c21)} The paying price is the highest bid from competitors, also called market price and auction winning price. If this bid price is higher than the auction winning price, then this record will occur in impression log. \\
\textsc{(c24)} User tags (segments) in iPinYou's proprietary audience database. Only a part of the user tags are released in this dataset.

\subsection{Basic Statistics}\label{sec:basic-stats}
The advertisers\footnote{Every advertiser of this dataset has only one campaign. Thus, these two terms are equivalent in this scenario.} and their industrial categories are summarised in Table~\ref{tab:adv-fields}. Note that in the 1st season no advertiser ID was given.
The diversity of advertisers makes the dataset more interesting.
As we show later in the paper, ads from different fields have greatly different user response behaviour.

\begin{table}[t]
\center
\caption{Advertiser Fields}\label{tab:adv-fields}
\vspace{0px}
\begin{tabular}{c c l}
Advertiser ID & Season & Industrial Category\\ \hline
\\[-2.0ex]
1458 & 2 & Chinese vertical e-commerce\\
2259 & 3 & Milk powder\\
2261 & 3 & Telecom\\
2821 & 3 & Footwear\\
2997 & 3 & Mobile e-commerce app install\\
3358 & 2 & Software\\
3386 & 2 & International e-commerce\\
3427 & 2 & Oil\\
3476 & 2 & Tire
\end{tabular}
\end{table}

\begin{table*}[t]
\center
\caption{Dataset statistics}\label{tab:basic-stats}
\vspace{5px}
\small
\begin{tabular}{rrrrrrrrrrrr}
\multicolumn{12}{c}{\textbf{Training Data}} \vspace{5px} \\
Adv. & Period~ & Bids & Imps & Clicks & Convs & Cost & Win Ratio & CTR & CVR & CPM & eCPC\\ \hline
\\[-2.0ex]
1458 & 6-12 Jun. & 14,701,496 & 3,083,056 & 2,454 & 1 & 212,400 & 20.97\% & 0.080\% & 0.041\% & 68.89 & 86.55\\
2259 & 19-22 Oct. & 2,987,731 & 835,556 & 280 & 89 & 77,754 & 27.97\% & 0.034\% & 31.786\% & 93.06 & 277.70\\
2261 & 24-27 Oct. & 2,159,708 & 687,617 & 207 & 0 & 61,610 & 31.84\% & 0.030\% & 0.000\% & 89.60 & 297.64\\
2821 & 21-23 Oct. & 5,292,053 & 1,322,561 & 843 & 450 & 118,082 & 24.99\% & 0.064\% & 53.381\% & 89.28 & 140.07\\
2997 & 23-26 Oct. & 1,017,927 & 312,437 & 1,386 & 0 & 19,689 & 30.69\% & 0.444\% & 0.000\% & 63.02 & 14.21\\
3358 & 6-12 Jun. & 3,751,016 & 1,742,104 & 1,358 & 369 & 160,943 & 46.44\% & 0.078\% & 27.172\% & 92.38 & 118.51\\
3386 & 6-12 Jun. & 14,091,931 & 2,847,802 & 2,076 & 0 & 219,066 & 20.21\% & 0.073\% & 0.000\% & 76.92 & 105.52\\
3427 & 6-12 Jun. & 14,032,619 & 2,593,765 & 1,926 & 0 & 210,239 & 18.48\% & 0.074\% & 0.000\% & 81.06 & 109.16\\
3476 & 6-12 Jun. & 6,712,268 & 1,970,360 & 1,027 & 26 & 156,088 & 29.35\% & 0.052\% & 2.532\% & 79.22 & 151.98\\
Total & - & 64,746,749 & 15,395,258 & 11,557 & 935 & 1,235,875 & 23.78\% & 0.075\% & 8.090\% & 80.28 & 106.94\\
\end{tabular}
\begin{tabular}{rrrrrrrrrrr}
\multicolumn{11}{c}{}\\
\multicolumn{11}{c}{\textbf{Test Data}} \vspace{5px} \\
Adv. & Period~ & Imps & Clicks & Convs & Cost & CTR & CVR & CPM & eCPC & $N$ \\ \hline
\\[-2.0ex]
1458 & 13-15 Jun. & 614,638 & 543 & 0 & 45,216 & 0.088\% & 0.000\% & 73.57 & 83.27 & 0 \\
2259 & 22-25 Oct. & 417,197 & 131 & 32 & 43,497 & 0.031\% & 24.427\% & 104.26 & 332.04 & 1 \\
2261 & 27-28 Oct. & 343,862 & 97 & 0 & 28,795 & 0.028\% & 0.000\% & 83.74 & 296.87 & 0 \\
2821 & 23-26 Oct. & 661,964 & 394 & 217 & 68,257 & 0.060\% & 55.076\% & 103.11 & 173.24 & 1 \\
2997 & 26-27 Oct. & 156,063 & 533 & 0 & 8,617 & 0.342\% & 0.000\% & 55.22 & 16.17 & 0 \\
3358 & 13-15 Jun. & 300,928 & 339 & 58 & 34,159 & 0.113\% & 17.109\% & 113.51 & 100.77 & 2 \\
3386 & 13-15 Jun. & 545,421 & 496 & 0 & 45,715 & 0.091\% & 0.000\% & 83.82 & 92.17 & 0 \\
3427 & 13-15 Jun. & 536,795 & 395 & 0 & 46,356 & 0.074\% & 0.000\% & 86.36 & 117.36 & 0 \\
3476 & 13-15 Jun. & 523,848 & 302 & 11 & 43,627 & 0.058\% & 3.642\% & 83.28 & 144.46 & 10 \\
Total & - & 4,100,716 & 3,230 & 318 & 364,243 & 0.079\% & 9.845\% & 88.82 & 112.77 & - \\
\end{tabular}
\end{table*}

The basic statistical information is given in Table~\ref{tab:basic-stats}. Specifically, the ``Win Ratio" column is about the ad auction winning ratio with the default bidding strategy from iPinYou platform. Conversion rate (CVR) is with respect to the number of clicks (instead of impressions). Note that in the original record, there would be multiple clicks on the same impression. However, duplications are removed in our analysis to allow focus on the events themselves (whether users would click or convert, or not).

From Table~\ref{tab:basic-stats} we can see that (i) all the advertisers has CTR less than 0.1\% except for advertiser 2997 (0.444\%). Note that 0.1\% is usually around the average CTR for desktop display advertising in practice. The high CTR for advertiser 2997 confirms the difference of mobile environment where clicks are more easily generated possibly due to ``fat finger'' effect; (ii) Although the nine advertisers have similar CPM, their effective cost-per-click (eCPC), i.e. the expected cost for achieving one click, are fairly different. This could be caused by the target rule setting (i.e., the target user demographic information, location and time) and the market of each specific advertiser; (iii) Some advertisers do not record conversions. Even for the ones who did report conversions, their CVRs differ a lot, which could also be due to different market and conversion setting.
In the table of test data, there is a conversion weight factor for each advertiser, denoted by $N$. It shows the relative importance of a conversion against a click for each advertiser. For example, the weight factor for advertiser 3476 (tire, $N=10$) is much higher than that for advertiser 2259 (milk powder, $N=1$).


\subsection{User Feedback}\label{sec:ctr}

\begin{figure}[t]
 \centering
 \includegraphics[width=\columnwidth]{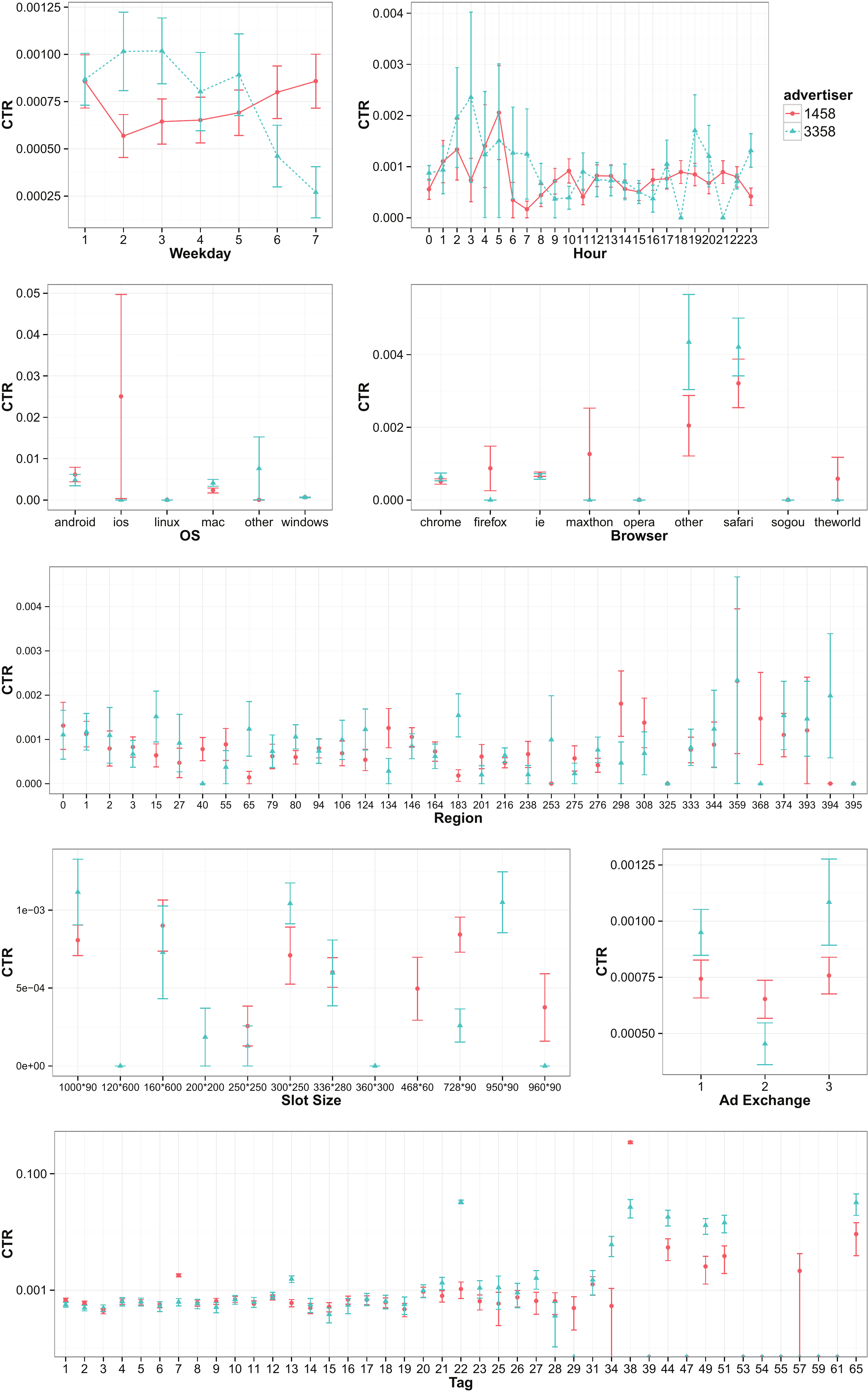}
 \caption{CTR distribution against different features for advertiser 1458 and 3358. We choose only two advertisers here because of the page limit and presentation concern.}
 \label{fig:ctr-1458-3358}
\end{figure}

Figure~\ref{fig:ctr-1458-3358} depicts some statistics of user feedback on advertiser 1458 and 3358. Specifically, the mean value with the standard error of CTR\footnote{We do not compare the CVR here because the conversion definitions across different advertisers are very different. For example, advertiser 2821 sells footwear and the conversion is defined as a purchase, while advertiser 3358 sells software and the conversion is a download.} against some features, such as the time, location, user-agent, publisher's ad slot size, ad exchanges, and user tags\footnote{Tags are re-indexed by the descending rank of their frequency.}.

We can see from Figure~\ref{fig:ctr-1458-3358} that for different advertisers, the same feature could have a different impact on the  CTR: \\
\textsc{(i)} Advertiser 1458 has received the highest CTR on Monday and weekends while advertiser 3358 does on Tuesday and Wednesday. \\
\textsc{(ii)} The mobile users (on Andriod or iOS) are more likely to click the ads from Advertiser 1458 while PC users (on Mac and Windows) prefer the ads from Advertiser 3358. \\
\textsc{(iii)} The ad CTR from two advertisers are both volatile across different region locations, and the trend are different. \\
\textsc{(iv)} Ad slot size is correlated with the slot locations in the webpage and the design of creatives. We can see the banner ($1000\times 90$) and standard ($300\times 250$) slots generally have the highest CTR for both advertisers. \\
\textsc{(v)} Ad exchanges call for bids and host the auctions. Different publishers (or their supply-side platforms, i.e. SSPs) connect different exchanges, thus the CTR distribution on these exchanges are different. \\
\textsc{(vi)} The CTR against different user tags is depicted in a log scale because the difference is fairly large. For example, the vertical e-commerce advertiser 1458 could receive the CTR as high as 30\% on the users with the tag 38 (\texttt{In-market/clothing, shoes\&bags}) while only around 0.1\% CTR on the other users. The same volatility happens to advertiser 3358. It shows the importance of user segmentation for predicting their response for a specific ad campaign. Therefore, the advertisers can refine their targeting rules and bidding strategies based on the ad performance on different user segmentations. Such user segmentation data is often provided by a third-party data management platform (DMP) or DSPs themselves.

In sum, the above analysis suggests that the user response models need to be trained independently for each advertiser. It requires some non-trivial work \cite{ahmed2014scalable} to leverage data from the similar advertisers to improve the performance of prediction. In addition, advertisers may not allow the DSP to use their data to help other advertisers.

\subsection{Bidding Behaviour}\label{sec:market-price}
In the second price auctions, the second highest bids are defined as the \emph{market price} for the winner. If his/her bid is higher than the market price, the advertiser wins this auction and pays the market price. Market price is always modelled as a stochastic variable because it is almost impossible to analyse the strategy of each of the thousands of auction participators \cite{amin2012budget}. A higher market price reflects a more competitive the environment.

\begin{figure}[t]
 \centering
 \includegraphics[width=\columnwidth]{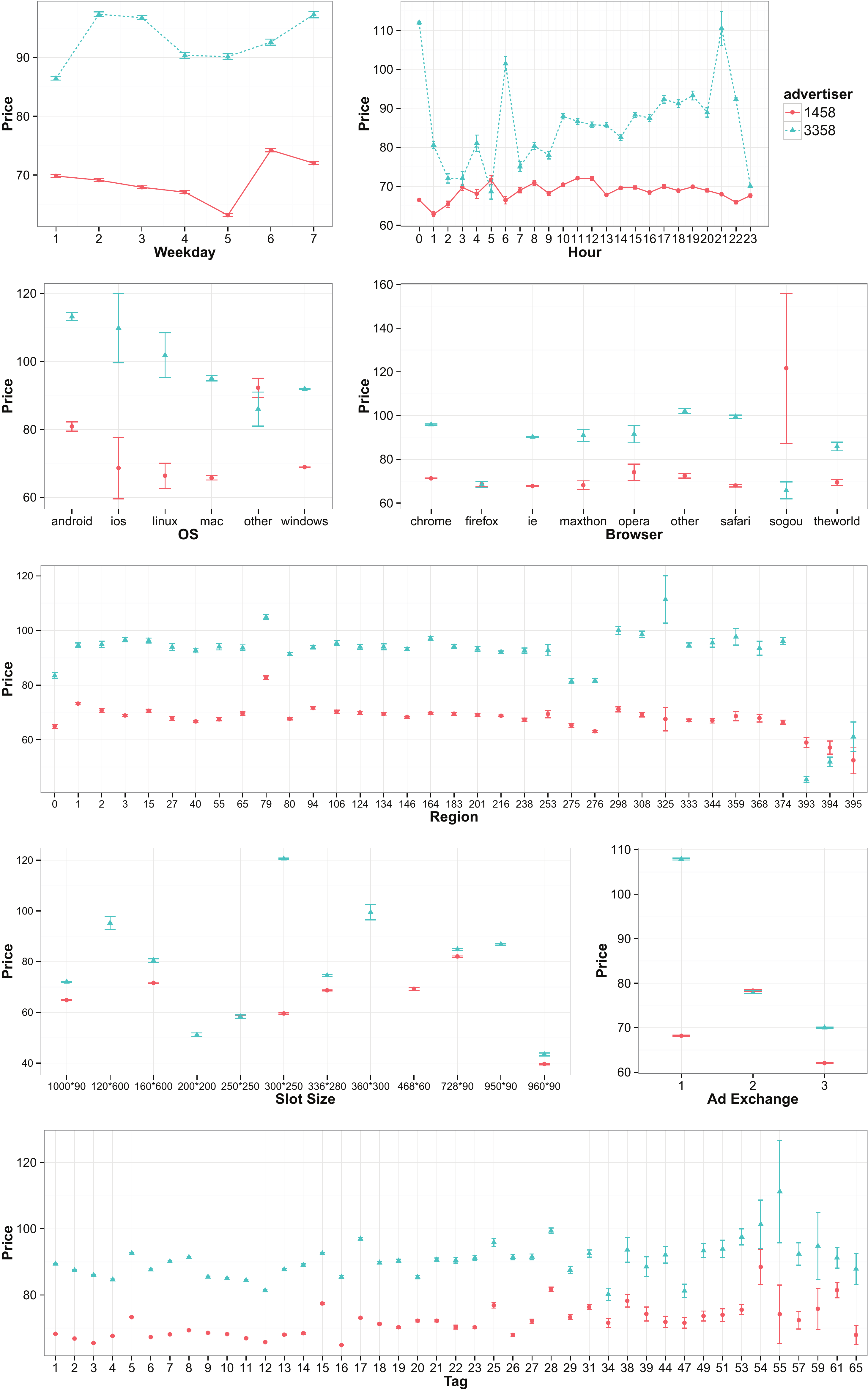}
 \caption{Market price distribution against different features for advertiser 1458 and 3358.}
 \label{fig:price-1458-3358}
\end{figure}

Here we have an investigation of the market price of advertiser 1458 and 3358 from the perspective of a DSP. The market price mean and standard error against different features are depicted in Figure~\ref{fig:price-1458-3358}, where we can see that just like the CTR plot, the market price has different trends against the same domain features on these two advertisers. For example, for advertiser 1458 the bid competitiveness in the morning is higher than that in the afternoon and evening, while it is inverse for advertiser 3358. In addition, for advertiser 1458 the competitiveness in ad exchange 1 is higher than that in ad exchange 2 and 3, while for advertiser 3358, the ad exchange 2 is the most competitive one.

Comparing the market price and CTR distribution on individual features, we find that the ratio of standard error to its mean of market prices is smaller than that of CTR. This is mainly because the click observations are binary value while market prices are integers.

\subsection{eCPC}\label{sec:ecpc}

\begin{figure}[t]
 \centering
 \includegraphics[width=\columnwidth]{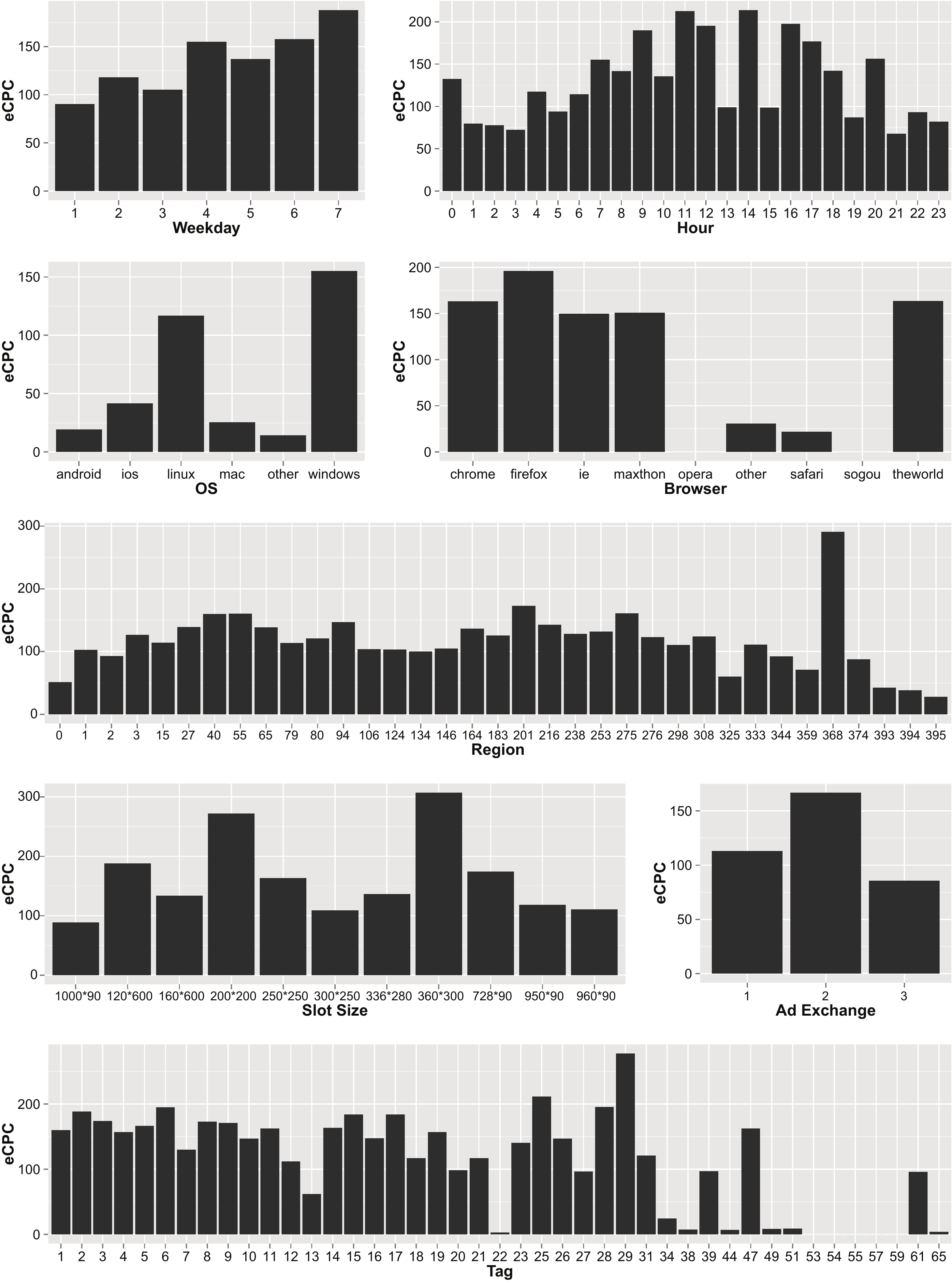}
 \caption{eCPC against different features for advertiser 3358. Zero-height bars mean there is no observed click for the specific feature.}
 \label{fig:ecpc-3358}
\end{figure}

From the joint observation on Figure~\ref{fig:ctr-1458-3358} and \ref{fig:price-1458-3358}, if we regard user clicks as the return, we can find some imbalanced return-on-investment (ROI) across the cases with different features. For example, we consider the weekday features for advertiser 3358. It has the lowest CTR but the highest market price on Sunday. Compared with Sunday, it has a lower market price but an around three-time CTR on Thursday. Another example for advertiser 1458 is that it has a quite lower market price on ad exchange 3 than that on ad exchange 2, but a higher CTR there. Such ROI is effectively measured by eCPC, which shows the amount of money that needs to be spent to achieve one click. A lower eCPC suggests a more cost effective algorithm.

We depict the eCPC bars against different features for advertiser 3358 in Figure~\ref{fig:ecpc-3358}. Just as the first example above, advertiser 3358 suffers the highest eCPC on Sunday, while it is much cost effective on Monday and Wednesday. In fact, the eCPC varies greatly against almost every feature considered here. For example, for advertiser 3358, the advertising to iOS users is about 3 times cost effective than that to Windows users. Its two low ROI creatives have the size of $200\times200$ and $360\times300$. The auctions from ad exchange 3 are much more cost effective than those from ad exchange 2. The users with tag 22, 38, 44, 49, 51, and 65 show significantly higher interest on this advertiser's ads than other users.

Ideally, if certain kind of features brings a lower eCPC than average, the advertiser (or DSP) should allocate more budget (via bidding higher) in such auctions. The observation from Figure~\ref{fig:ecpc-3358} indicates that there is great optimisation potential. For example, if the advertiser reallocates part of the budget from Sunday to Monday, more clicks could be achieved with the same budget. But we cannot bid much higher on Monday, because the higher bid price results in higher cost, which always increases the eCPC. Thus, there is a trade-off between the achieved clicks and the eCPC. Due to the page limit, we do not show the eCPC performance for other advertisers. In fact, all advertisers have the inconsistent eCPC across different features, and the changing trends across the features are different, too.

From the above data analysis we can see the same features would have much different impact on the user feedback and market price of different advertisers, which results in different eCPC. Therefore, it is reasonable to independently build the user response models and bidding strategies for each advertiser.

\section{Task and Protocol}\label{sec:task}

\begin{figure*}
 \vspace{-0pt}
 \centering
 \includegraphics[width=0.9\textwidth]{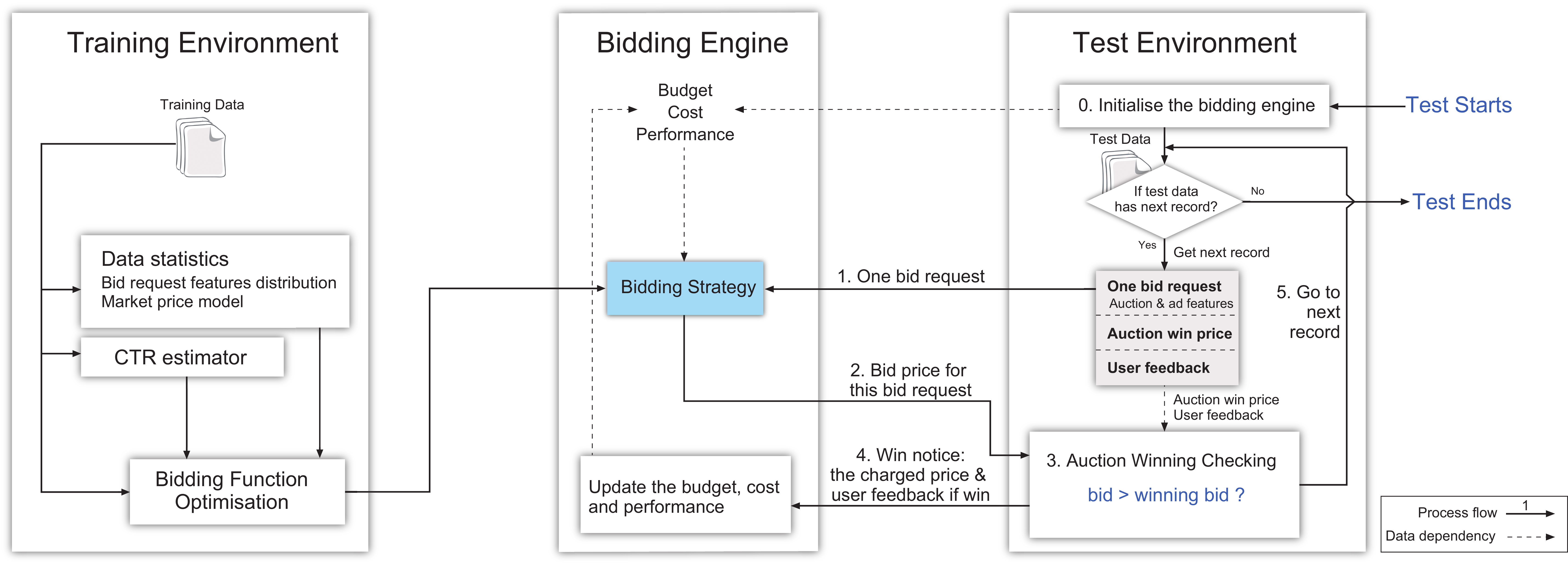}
 \caption{The training framework and evaluation protocol.}
 \label{fig:training-evaluation}
\end{figure*}

\subsection{Task Description}\label{sec:task}
The DSP bid optimisation task for a given advertiser refers to optimising a predefined KPI given a cost budget and the coming bid requests during the lifetime of the budget. For example, a straightforward KPI is the click (conversion) number and the task is to maximise the click (conversion) number with the given budget and coming bid requests. We write the general optimisation formula as:
\begin{align}
\max_{\text{bidding strategy}} &~~~ \text{KPI}\\
\text{subject to} &~~~ \text{cost} \leq \text{budget}.
\end{align}

In the iPinYou bidding algorithm competition, the KPI was a linear combination of the click number and conversion number:
\begin{align}
\max_{\text{bidding strategy}} &~~~ \text{\#click} + N \cdot \text{\#conversion} \label{eq:ipinyou-kpi} \\
\text{subject to} &~~~ \text{cost} \leq \text{budget},
\end{align}
where $N$ is the parameter showing the relative importance of conversions to clicks and it varies across different advertisers, as has been shown in Table~\ref{tab:basic-stats}. This KPI is practically meaningful since conversions are the final measure for advertising but they are usually sparse.

As shown in Figure~\ref{fig:interaction}, the input of the bidding strategy is a bid request (with the auction and ad information), the output is the bid response for this bid request. In addition, in the process of making a bid decision, the budget, current cost and achieved performance is accessible to the bidding engine.

Figure~\ref{fig:training-evaluation} presents a flow diagram about the bidding strategy training framework and the following test evaluation protocol. In the latter two subsections, we discuss these two parts respectively.

\subsection{Training Framework}\label{sec:training-framework}
As discussed in Section~\ref{sec:dataset}, given the training data, one can investigate the data to learn about the bid request feature distribution and derive its relationship with the user ad response (e.g., CTR) and the market price. As shown in the left part of Figure~\ref{fig:training-evaluation}, based on the components of the CTR estimator and the market price model, one can use the training data to perform the bidding function optimisation and finally obtain the bidding strategy.

With the feature engineering on the bid request data and the label extraction from user feedback data, one can train a CTR estimator to predict the probability of the user click on a given ad impression. Standard regression models such Logistic regression and tree models can be used here. We will later show the details in the experiment section.

Besides the CTR estimation, the market price distribution can also be estimated either by direct data statistics or a regression model, which is called bid landscape forecasting \cite{cui2011bid}. With the knowledge of the market price distribution, one can estimate the probability of winning a specific ad auction and the corresponding cost given a bid price.

A bidding function generally takes the predicted CTR (pCTR) and the market price distribution as input and outputs the bid price. As a framework description, here we do not specify the process of optimising the bidding function, which could be different for different models. In fact, the market price model has not been formally considered in previous bid optimisation work \cite{perlich2012bid,lee2012estimating}. In addition, other factors can also be added into this training framework.

\subsection{Evaluation Protocol}\label{sec:eval-protocol}

The evaluation protocol is illustrated as in the middle and right part of Figure~\ref{fig:training-evaluation}.  Given the bidding strategy and a budget for the test period for a particular advertiser, we can perform a simulation via going through its bid logs and comparing the bidding results with impression, click, and conversion logs. A detailed description of each step is as follow: \\
\textsc{(0)} To initialise the bidding engine, such as setting a predefined budget, initialising the cost and performance (e.g., achieved click and conversion number) as zero.\\
\textsc{(1)} To pass the next bid request (in the ascending order of the timestamp) to the bidding engine. A bid request contains both contextual and behavioural data of the auction as well as the ad data as shown in Table~\ref{tab:data-format}. \\
\textsc{(2)} The bidding strategy computes a bid for this request (with the information of the budget, current cost and achieved performance). This step is highlighted in Figure~\ref{fig:training-evaluation} as it is what bid optimisation focuses on. Note that if the cost has been higher than the budget (i.e. the budget is run out), all the bid responses should be set to zero (i.e. to skip all the left bid requests).\\
\textsc{(3)} To simulate the auction by referencing the impression logs: if the bid price is higher than the logged auction winning price (i.e., paying price on column 21 in Table~\ref{tab:data-format}) and floor price (column 18 in Table~\ref{tab:data-format}), the bidding engine wins the auction and gets the ad impression. \\
\textsc{(4)} To match the click and conversion events in the logs for this impression if winning the auction. The performance of the bidding strategy is then updated and saved. The cost is also added by the paying price. \\
\textsc{(5)} To check whether there is any bid request left in test data to determine if the evaluation needs to terminate. Note that a tiny amount of over-spend from the last impression is possible but it is neglected in our evaluation.

Compared the procedures in Figure~\ref{fig:interaction} and \ref{fig:training-evaluation}, we can see step 1,2 and 3 are the same. The step 4 in evaluation flow actually merges the step 4 (win notice), 5 (ad delivery and tracking) and 6 (user feedback notice) in Figure~\ref{fig:interaction}, which is reasonable for offline evaluation since the log data has already collected the user feedback information.

It is worth noting that there are limitations of replaying logs for offline evaluation. Click and conversion events are only available for winning auctions (having impressions); there is no data for the lost auctions. Thus, there is no way to check if the performance could be improved if the algorithm bids higher to win those lost auctions.
However, this evaluation protocol do follow the convention of the offline evaluations from sponsored search \cite{zhang2012joint,graepel2010web}, recommender systems \cite{zhang2013optimizing} and Web search \cite{craswell2008experimental} where the objects (auctions) with unseen user feedback have to be ignored.

\section{Benchmark Experiments}
As discussed before, it is routine to perform the real-time bidding with a two stage process \cite{perlich2012bid,lee2012estimating}: (i) estimating the CTR/CVR of the ad impression being auctioned; (ii) making the bid decision based on the evaluation and other information. For iPinYou dataset, different advertisers have largely different setting on their conversions and more than half of them has no conversion record. Therefore, in our benchmark experiment, we will focus on the clicks. We first test the performance of two standard CTR estimators. Then we compared several bidding strategies based on the pCTR. Besides the total achieved clicks, the considered KPIs include conversions and the iPinYou KPI in Eq. (\ref{eq:ipinyou-kpi}).

\subsection{CTR Estimation}
\subsubsection{Compared Models and Feature Engineering}
In our benchmarking, we consider the following two CTR estimation models. Besides the model setting, the feature engineering details are also discussed respectively.

\noindent \textbf{Logistic Regression (\textsf{LR})} is a widely used linear model to estimate CTR in computational advertising \cite{Richardson2007a}. The loss is the cross entropy between the predicted click probability and the ground-truth result. In addition, L2 regularisation is used.

In our experiment, all the features for \textsf{LR} are binary. Specifically, we extract the weekday and hour feature from timestamps. User agent text is processed to extract the operation systems and browser brands as shown in Figure~\ref{fig:ctr-1458-3358}. The floor price is processed by buckets of 0, [1,10], [11,50], [51,100] and [101,+$\infty$). The tag list of each user is divided into binary features for each tag. We do not include the features of Bid ID, Log Type, iPinYou ID, URL, Anonymous URL ID, Bidding Price, Paying Price, Key Page URL because they are either almost unique for each case or meaningless to be added to \textsf{LR} training. Also we do not add combination features (e.g., weekday-region-tag) because there are many variants and tricks for adding high-order combination features, which is not recommended in benchmark experiment. In sum, we have 937,748 binary features for \textsf{LR} training and prediction.

\noindent \textbf{Gradient Boosting Regression Tree (\textsf{GBRT})} \cite{friedman2002stochastic} is a non-linear model widely used in regression and learning to rank applications \cite{burges2010ranknet}. Comparing to the linear model \textsf{LR}, \textsf{GBRT} has the advantage of learning the non-linear features, which can hardly be achieved by the feature engineering of \textsf{LR}. However, once finishing training, \textsf{GBRT} will get rid of most features and only keep a small part of features for prediction. In our experiment, we leverage the open-sourced \textsc{xgboost}\footnote{https://github.com/tqchen/xgboost} for implementation. Specifically, we set the max tree depth as 5 and train 50 trees with 0.05 learning rate.

Different with the binary features for \textsf{LR}, here every feature for \textsf{GBRT} is continuous. Specifically, for the indicator features (e.g., city=152) of a domain (e.g., city) we calculate the frequency and CTR based on a subset of the training file and let these two numbers be the values of two \textsf{GBRT} features (e.g., the frequency that city=152 is 348,432, the empirical CTR for the cases whose city=152 is 0.1\%). We do not make the frequency value for tag features. For the continuous feature (e.g., slot floor price), we directly use the specific value as the feature value (e.g., slot floor price=21). In sum, we have 78 features for \textsf{GBRT} training and prediction.

\subsubsection{Evaluation Measure}
The area under ROC curve (AUC) is a widely used measure for evaluating the ad CTR estimator \cite{graepel2010web,oentaryo2014predicting}. Besides AUC, the root mean square error (RMSE) is also chosen as the evaluation measure here as it is widely used in various regression tasks. Because of the huge imbalance of positive/negative cases in ad clicking, the empirically best regression model usually provides the pCTR very close to 0, which results in the RMSE having a quite small value and the improvement on RMSE is much slight, compared with AUC.

\begin{table}[t]
\center
\caption{CTR estimation performance.}
\label{tab:ctr-estimation}
\begin{tabular}{cc|cc|cc}
\small
 & & \multicolumn{2}{c|}{AUC} & \multicolumn{2}{c}{RMSE}\\
Season & Adv. & \textsf{LR} & \textsf{GBRT} & \textsf{LR} & \textsf{GBRT}\\
\hline
2 & 1458 & 0.9881 & 0.9707 & 0.0191 & 0.0263\\
3 & 2259 & 0.6865 & 0.6791 & 0.0177 & 0.0176\\
3 & 2261 & 0.6238 & 0.5739 & 0.0168 & 0.0167\\
3 & 2821 & 0.6325 & 0.5820 & 0.0239 & 0.0238\\
3 & 2997 & 0.6039 & 0.5979 & 0.0582 & 0.0581\\
2 & 3358 & 0.9753 & 0.9722 & 0.0246 & 0.0279\\
2 & 3386 & 0.7908 & 0.7686 & 0.0284 & 0.0285\\
2 & 3427 & 0.9735 & 0.9342 & 0.0214 & 0.0245\\
2 & 3476 & 0.9625 & 0.9422 & 0.0230 & 0.0231\\
2 & Total & 0.9141 & 0.9200 & 0.0263 & 0.0260\\
3 & Total & 0.7615 & 0.7715 & 0.0268 & 0.0268\\
2,3 & Total & 0.8307 & 0.8518 & 0.0270 & 0.0263\\
\end{tabular}
\end{table}

\subsubsection{Results}
The experimental results for CTR estimation with \textsf{LR} and \textsf{GBRT} are shown in Table~\ref{tab:ctr-estimation}. From the results we can see different advertisers have quite large difference on the value of AUC and RMSE, due to the different user behaviour on their ads. For example, advertiser 2997 has the highest overall CTR (0.444\%) but the lowest observation number (see Table~\ref{tab:basic-stats}) which makes it more difficult to predict the CTR. In addition, both models achieve a much better performance on advertisers in season 2 than that in season 3. iPinYou technicians explained that this is due to the different user segmentation systems between season 2 and 3.

\subsection{DSP Bid Optimisation}
\subsubsection{Compared Bidding Strategies}
We compare the following bidding strategies in our benchmark experiment for DSP real-time bidding. The parameters of each bidding strategy are tuned using the training data. And the evaluation is performed on the test data.

\noindent \textbf{Constant bidding (\textsf{Const}).} Bid a constant value for all the bid requests. The parameter is the specific constant bid price.

\noindent \textbf{Random bidding (\textsf{Rand}).} Randomly choose a bid value in a given range. The parameter is the upper bound of the random bidding range.

\noindent \textbf{Bidding below max eCPC (\textsf{Mcpc}).} The goal of bid optimisation is to reduce the eCPC. In \cite{lee2012estimating}, given the advertiser's goal on max eCPC, which is the upper bound of expected cost per click, the bid price on an impression is obtained by multiplying the max eCPC and the pCTR. Here we calculate the max eCPC for each campaign by dividing its cost and achieved number of clicks in the training data. No parameter for this bidding strategy.

\noindent \textbf{Linear-form bidding of pCTR (\textsf{Lin}).} In the previous work \cite{perlich2012bid}, the bid value is linearly proportional to the pCTR under the target rules. The formula can be generally written as \texttt{bid=base\_bid$\times$pCTR/avgCTR}, where the tuning parameter \texttt{base\_bid} is the bid price for the average CTR cases.

Among the compared bidding strategies, only \textsf{Mcpc} is non-parametric and it does not consider the budget limits while the others do by tuning their parameters. In addition, \textsf{Mcpc} and \textsf{Lin} need to evaluate CTR for each impression. Thus we denote the two bidding strategies with \textsf{LR} CTR estimator as \textsf{Mcpc-L}, \textsf{Lin-L} and with \textsf{GBRT} CTR estimator as \textsf{Mcpc-G} and \textsf{Lin-G}.

\subsubsection{Experimental Setting}
The evaluation follows the protocol in Section~\ref{sec:eval-protocol}. The only issue discussed here is about the pre-set budget for each advertiser. In our experiment we set the budget for each advertiser as a proportion of the original total cost in the test log. Particularly, in order to check the bidding strategies' performance under different budget limits, we set the budget as 1/32, 1/8, and 1/2 of the original total cost in the test log. Note that we cannot set the budget higher than the original total cost because in such case, simply bidding as high as possible on each auction will make the DSP win all the auctions without running out of the budget.

\subsubsection{Results}
We list the achieved click performance for each algorithm under different budget limits in Table~\ref{tab:clk-perf}. We can see from Table~\ref{tab:clk-perf} that \textsf{Lin} and \textsf{Mcpc} generally work much better than \textsf{Const} and \textsf{Rand}, which verifies the importance of impression-level evaluation and real-time bidding. To some advertisers, such as 2259 and 2261 with 1/32 and 1/8 budget limits, \textsf{Mcpc} achieves fewer clicks than \textsf{Const} and \textsf{Rand}. This is because \textsf{Mcpc} is not adaptive with different budget limits, which results in running out of budget quite soon for the low budget settings. Moreover, \textsf{Lin} works much better than \textsf{Mcpc}. Compared with \textsf{Mcpc}, \textsf{Lin} has the ability to change the bidding scale by tuning its parameter \texttt{base\_bid}, which helps \textsf{Lin} adapt different budget limits against the coming ad auction volume. Such adaptivity is essential to DSP bidding strategies because the ad auction volume and market price could vary a lot as time goes by \cite{cui2011bid}. Generally, if the budget allocated per ad auction is high, then it is encouraged to bid high to achieve more user clicks and conversions, while if the budget allocated per ad auction is low, the bid price should be reduced to remain the high ROI.

\begin{table}[t]
\center
\vspace{-0pt}
\caption{Click numbers for each advertiser under different budget limits.}
\label{tab:clk-perf}
\small
\begin{tabular}{r|rrrrrr}
\multicolumn{7}{c}{\textbf{Budget (1/32)}}\\
Adv. & \textsf{Const} & \textsf{Rand} & \textsf{Mcpc-L} & \textsf{Mcpc-G} & \textsf{Lin-L} & \textsf{Lin-G}\\
\hline
\\[-2.0ex]
1458 & 28 & 29 & 261 & 89 & 500 & 491\\
2259 & 12 & 11 & 7 & 6 & 16 & 15\\
2261 & 9 & 9 & 3 & 3 & 12 & 11\\
2821 & 37 & 44 & 16 & 16 & 57 & 42\\
2997 & 74 & 63 & 22 & 46 & 78 & 78\\
3358 & 11 & 13 & 85 & 83 & 278 & 260\\
3386 & 23 & 23 & 39 & 29 & 127 & 86\\
3427 & 21 & 21 & 67 & 43 & 321 & 294\\
3476 & 27 & 25 & 33 & 28 & 205 & 140\\
S2 & 110 & 111 & 485 & 272 & 1431 & 1271\\
S3 & 132 & 127 & 48 & 71 & 163 & 146\\
Total & 242 & 238 & 533 & 343 & 1594 & 1417\\
\multicolumn{7}{c}{}\\
\multicolumn{7}{c}{\textbf{Budget (1/8)}}\\
Adv. & \textsf{Const} & \textsf{Rand} & \textsf{Mcpc-L} & \textsf{Mcpc-G} & \textsf{Lin-L} & \textsf{Lin-G}\\
\hline
\\[-2.0ex]
1458 & 92 & 98 & 502 & 294 & 521 & 496\\
2259 & 28 & 27 & 25 & 27 & 37 & 32\\
2261 & 25 & 26 & 23 & 20 & 28 & 25\\
2821 & 88 & 88 & 61 & 84 & 109 & 91\\
2997 & 159 & 138 & 78 & 112 & 162 & 166\\
3358 & 50 & 59 & 310 & 272 & 314 & 284\\
3386 & 67 & 72 & 128 & 112 & 216 & 184\\
3427 & 56 & 59 & 361 & 205 & 358 & 321\\
3476 & 63 & 71 & 148 & 94 & 273 & 235\\
S2 & 328 & 359 & 1449 & 977 & 1682 & 1520\\
S3 & 300 & 279 & 187 & 243 & 336 & 314\\
Total & 628 & 638 & 1636 & 1220 & 2018 & 1834\\
\multicolumn{7}{c}{}\\
\multicolumn{7}{c}{\textbf{Budget (1/2)}}\\
Adv. & \textsf{Const} & \textsf{Rand} & \textsf{Mcpc-L} & \textsf{Mcpc-G} & \textsf{Lin-L} & \textsf{Lin-G}\\
\hline
\\[-2.0ex]
1458 & 322 & 315 & 502 & 503 & 540 & 524\\
2259 & 73 & 79 & 86 & 86 & 96 & 87\\
2261 & 66 & 67 & 70 & 66 & 72 & 67\\
2821 & 240 & 236 & 239 & 124 & 245 & 253\\
2997 & 366 & 355 & 329 & 112 & 358 & 359\\
3358 & 163 & 174 & 310 & 272 & 335 & 307\\
3386 & 276 & 277 & 336 & 280 & 401 & 369\\
3427 & 208 & 217 & 361 & 329 & 389 & 363\\
3476 & 195 & 186 & 290 & 274 & 301 & 279\\
S2 & 1164 & 1169 & 1799 & 1658 & 1966 & 1842\\
S3 & 745 & 737 & 724 & 388 & 771 & 766\\
Total & 1909 & 1906 & 2523 & 2046 & 2737 & 2608\\
\end{tabular}
\end{table}

Table~\ref{tab:cnv-perf} lists the conversion performance on four advertisers with conversion records. And Table~\ref{tab:score-perf} lists the corresponding iPinYou KPI score of $\text{\#click}+N\cdot\text{\#conversion}$, where $N$ for each advertiser has been shown in Table~\ref{tab:score-perf}. We can see the significant improvement of \textsf{Lin} against other bidding strategies. Another common important point from the results of these three KPIs is that in the lower budget setting, \textsf{Lin} achieves the higher improvement rate against other strategies. This is because when the budget is quite limited, it is more important to identify which cases are probably valuable and adaptively lower the overall bid price.

\begin{table}[t]
\center
\caption{Conversion numbers for each advertiser under different budget limits.}
\label{tab:cnv-perf}
\small
\begin{tabular}{r|rrrrrr}
\multicolumn{7}{c}{\textbf{Budget (1/32)}}\\
Adv. & \textsf{Const} & \textsf{Rand} & \textsf{Mcpc-L} & \textsf{Mcpc-G} & \textsf{Lin-L} & \textsf{Lin-G}\\
\hline
\\[-2.0ex]
2259 & 4 & 4 & 3 & 4 & 4 & 4\\
3476 & 1 & 2 & 1 & 1 & 3 & 3\\
2821 & 24 & 29 & 7 & 8 & 33 & 29\\
3358 & 4 & 4 & 14 & 14 & 53 & 51\\
S2 & 5 & 6 & 15 & 15 & 56 & 54\\
S3 & 28 & 33 & 10 & 12 & 37 & 33\\
Total & 33 & 39 & 25 & 27 & 93 & 87\\
\multicolumn{7}{c}{}\\
\multicolumn{7}{c}{\textbf{Budget (1/8)}}\\
Adv. & \textsf{Const} & \textsf{Rand} & \textsf{Mcpc-L} & \textsf{Mcpc-G} & \textsf{Lin-L} & \textsf{Lin-G}\\
\hline
\\[-2.0ex]
2259 & 7 & 9 & 7 & 8 & 10 & 10\\
3476 & 3 & 3 & 4 & 3 & 8 & 5\\
2821 & 54 & 51 & 35 & 49 & 61 & 54\\
3358 & 10 & 11 & 56 & 52 & 57 & 53\\
S2 & 13 & 14 & 60 & 55 & 65 & 58\\
S3 & 61 & 60 & 42 & 57 & 71 & 64\\
Total & 74 & 74 & 102 & 112 & 136 & 122\\
\multicolumn{7}{c}{}\\
\multicolumn{7}{c}{\textbf{Budget (1/2)}}\\
Adv. & \textsf{Const} & \textsf{Rand} & \textsf{Mcpc-L} & \textsf{Mcpc-G} & \textsf{Lin-L} & \textsf{Lin-G}\\
\hline
\\[-2.0ex]
2259 & 16 & 19 & 19 & 20 & 23 & 20\\
3476 & 6 & 8 & 9 & 7 & 11 & 7\\
2821 & 140 & 133 & 130 & 75 & 134 & 139\\
3358 & 28 & 31 & 56 & 52 & 58 & 57\\
S2 & 34 & 39 & 65 & 59 & 69 & 64\\
S3 & 156 & 152 & 149 & 95 & 157 & 159\\
Total & 190 & 191 & 214 & 154 & 226 & 223\\
\end{tabular}
\end{table}

\begin{table}[t]
\center
\caption{KPI score (\#clicks+$N\cdot$\#conversions) for each advertiser under different budget limits.}
\label{tab:score-perf}
\small
\begin{tabular}{r|rrrrrr}
\multicolumn{7}{c}{\textbf{Budget (1/32)}}\\
Adv. & \textsf{Const} & \textsf{Rand} & \textsf{Mcpc-L} & \textsf{Mcpc-G} & \textsf{Lin-L} & \textsf{Lin-G}\\
\hline
\\[-2.0ex]
1458 & 28 & 29 & 261 & 89 & 500 & 491\\
2259 & 16 & 15 & 10 & 10 & 20 & 19\\
2261 & 9 & 9 & 3 & 3 & 12 & 11\\
2821 & 61 & 73 & 23 & 24 & 90 & 71\\
2997 & 74 & 63 & 22 & 46 & 78 & 78\\
3358 & 19 & 21 & 113 & 111 & 384 & 362\\
3386 & 23 & 23 & 39 & 29 & 127 & 86\\
3427 & 21 & 21 & 67 & 43 & 321 & 294\\
3476 & 37 & 45 & 43 & 38 & 235 & 170\\
S2 & 128 & 139 & 523 & 310 & 1567 & 1403\\
S3 & 160 & 160 & 58 & 83 & 200 & 179\\
Total & 288 & 299 & 581 & 393 & 1767 & 1582\\
\multicolumn{7}{c}{}\\
\multicolumn{7}{c}{\textbf{Budget (1/8)}}\\
Adv. & \textsf{Const} & \textsf{Rand} & \textsf{Mcpc-L} & \textsf{Mcpc-G} & \textsf{Lin-L} & \textsf{Lin-G}\\
\hline
\\[-2.0ex]
1458 & 92 & 98 & 502 & 294 & 521 & 496\\
2259 & 35 & 36 & 32 & 35 & 47 & 42\\
2261 & 25 & 26 & 23 & 20 & 28 & 25\\
2821 & 142 & 139 & 96 & 133 & 170 & 145\\
2997 & 159 & 138 & 78 & 112 & 162 & 166\\
3358 & 70 & 81 & 422 & 376 & 428 & 390\\
3386 & 67 & 72 & 128 & 112 & 216 & 184\\
3427 & 56 & 59 & 361 & 205 & 358 & 321\\
3476 & 93 & 101 & 188 & 124 & 353 & 285\\
S2 & 378 & 411 & 1601 & 1111 & 1876 & 1676\\
S3 & 361 & 339 & 229 & 300 & 407 & 378\\
Total & 739 & 750 & 1830 & 1411 & 2283 & 2054\\
\multicolumn{7}{c}{}\\
\multicolumn{7}{c}{\textbf{Budget (1/2)}}\\
Adv. & \textsf{Const} & \textsf{Rand} & \textsf{Mcpc-L} & \textsf{Mcpc-G} & \textsf{Lin-L} & \textsf{Lin-G}\\
\hline
\\[-2.0ex]
1458 & 322 & 315 & 502 & 503 & 540 & 524\\
2259 & 89 & 98 & 105 & 106 & 119 & 107\\
2261 & 66 & 67 & 70 & 66 & 72 & 67\\
2821 & 380 & 369 & 369 & 199 & 379 & 392\\
2997 & 366 & 355 & 329 & 112 & 358 & 359\\
3358 & 219 & 236 & 422 & 376 & 451 & 421\\
3386 & 276 & 277 & 336 & 280 & 401 & 369\\
3427 & 208 & 217 & 361 & 329 & 389 & 363\\
3476 & 255 & 266 & 380 & 344 & 411 & 349\\
S2 & 1280 & 1311 & 2001 & 1832 & 2192 & 2026\\
S3 & 901 & 889 & 873 & 483 & 928 & 925\\
Total & 2181 & 2200 & 2874 & 2315 & 3120 & 2951\\
\end{tabular}
\end{table}

\section{Other Research Topics}
Besides the CTR estimation and DSP bid optimisation problem, there are other potential research topics where this dataset can be used for experimentation.
\begin{itemize}
\item \textbf{Bid landscape modelling.} As previously mentioned, for RTB based display advertising, market price can varying quite a lot \cite{cui2011bid}. The current DSP work \cite{perlich2012bid,lee2012estimating} mainly focuses on the bidding strategy based on CTR/CVR estimation. However, another important factor for the bidding decision making is the market price. As a repeated auction game with the budget constraint, the optimal bidding strategy is not truth-telling while it does depend on the market price distribution. Thus the research problem of modelling the market price distribution is much meaningful for a better bidding strategy.
\item \textbf{Adaptive bid control.} Just like sponsored search, a practically significant problem is to adaptively control the scale of real-time bidding to exactly deliver the budget during the campaign's lifetime. For some advertisers setting a high bidding scale at the beginning, their budget could be run out just at the beginning of the campaign lifetime. The techniques of pacing \cite{lee2013real} could be used here. More generally, in \cite{amin2012budget}, the authors model the bid control problem in sponsored search via Markov decision processes (MDPs) where the left budget and ad auction volume are states while the bid prices are actions. Further research on adaptive bid control can be based on this dataset.
\end{itemize}

\section{Conclusions}
Due to the sensitivity, research on computational advertising, especially from academia, is seriously restricted by the data availability. With the publication of iPinYou dataset, we believe the research on RTB display advertising will be stimulated.

In this paper, we performed a detailed statistical analysis on iPinYou dataset; formally defined the bid optimisation problem; and presented a simple yet comprehensive offline evaluation protocol for bidding strategies. We conducted benchmark experiments on CTR estimation and DSP bid optimisation with several baseline algorithms. The dataset and report are hosted on our website for computational advertising research.

{
\bibliographystyle{abbrv}
\bibliography{rtb-data}
}

\end{document}